\documentclass[twocolumn]{aa}
\usepackage{graphicx}
\usepackage{txfonts}
\begin{document}
\title{No Planet Around HD 219542 B
          \thanks{Based on observations made with the Italian Telescopio 
                  Nazionale Galileo (TNG) operated on the island of La Palma
                  by the Centro Galileo Galilei of the INAF (Istituto 
                  Nazionale di Astrofisica) at the Spanish Observatorio del 
                  Roque de los Muchachos of the Instituto de Astrofisica 
                  de Canarias}}

   \author{S. Desidera
          \inst{1},
           R.G. Gratton
          \inst{1}, 
           M. Endl
          \inst{2},
           R.U. Claudi
          \inst{1},
           R. Cosentino,
          \inst{3,4},
           M. Barbieri
          \inst{5,6},
           G. Bonanno
          \inst{3},
           S. Lucatello
          \inst{1,7},
           A.F. Martinez Fiorenzano
          \inst{1,7},
           F. Marzari
          \inst{8}
           \and 
           S. Scuderi
          \inst{2}}

   \authorrunning{S. Desidera et al.}

   \offprints{S. Desidera,  \\
              \email{desidera@pd.astro.it} }

   \institute{INAF -- Osservatorio Astronomico di Padova,  
              Vicolo dell' Osservatorio 5, I-35122, Padova, Italy 
             \and
             McDonald Observatory, The University of Texas at Austin, Austin, 
             TX 78712, USA         
             \and
             INAF -- Osservatorio Astrofisico di Catania, Via S.Sofia 78, Catania, Italy
             \and
             INAF -- Centro Galileo Galiei, Calle Alvarez de Abreu 70, 38700 
             Santa Cruz de La Palma (TF), Spain 
             \and
             CISAS -- Universit\`a di Padova, c/o Dipartimento di Fisica Via 
             Marzolo 8, Padova, Italy 
             \and
             LESIA, Observatoire de Paris, Section de Meudon, 92195 
             Meudon Principal Cedex, France
             \and
             Dipartimento di Astronomia -- Universit\`a di Padova, Vicolo
             dell'Osservatorio 2, Padova, Italy 
             \and 
             Dipartimento di Fisica -- Universit\`a di Padova, Via Marzolo 8,
             Padova, Italy }

 \date{}

   \abstract{The star HD~219542~B has been reported by us (Desidera et al.
             2003) to show low-amplitude radial velocity variations
             that could be due to the presence of a Saturn-mass planetary
             companion or to stellar activity phenomena.
             In this letter we present the results of the continuation of 
             the radial velocity monitoring as well as a discussion
             of literature determinations of the chromospheric activity of the 
             star (Wright et al.~2004).
             These new data indicate that the observed radial velocity
             variations are likely related to stellar activity. 
             In particular, there
             are indications that HD~219542~B underwent a phase of enhanced
             stellar activity in 2002 while the activity level has been
             lower in both 2001 and 2003. Our 2003 radial velocity 
             measurements now deviate from our preliminary orbital solution
             and the peak in the power spectrum at the proposed planet period 
             is severely reduced by the inclusion of the new data. 
             We therefore 
             dismiss the planet hypothesis as the cause of the radial velocity 
             variations.

   }

   \maketitle
%
%________________________________________________________________

\section{Introduction}
\label{s:intro}

In Desidera et al.~(\cite{hd219542}, hereafter Paper I) we presented 
high precision radial velocity (hereafter RV) monitoring of the 
components of the wide binary system HD~219542.
This pair is part of the sample of wide binaries
currently under monitoring at TNG using the high resolution spectrograph
SARG (Gratton et al.~\cite{papersarg}).
We have found evidence for the presence of low amplitude 
RV variations on the secondary HD~219542~B with a period of 112 days at a 
confidence level of 96-97\%.
These RV variations could be due to a Saturn-mass planet
orbiting at 0.46~AU or to stellar activity.

The relatively low statistical confidence of the proposed planetary 
orbit as well as the possible presence of stellar activity indicate that 
a confirmation is required before we can classify of HD~219542~B as a 
bona fide planet host star.

%%%%%%%%%%%%%%%%%%%%%%%%%%%%%%%%%%%%%%%%%%%%%%%%%%%%%%%%%%%%%%%%%

It is well known that stellar activity induces distortions of the profile 
of the spectral lines that could be seen as RV
variations and then mimic the occurrence of companions orbiting the target 
(see e.g. Saar et al.~\cite{saar}).
These spurious RV variations may have amplitudes and timescales comparable
to those induced by giant planets, making challenging the search for planets
around active stars. 
The controversial case of HD~192263 (Henry et al.~\cite{henry02}; 
Santos et al.~\cite{santos_hd192263}) is worth of mention in this context.

Two different components of the magnetic activity phenomena are of concern 
for planet searches.
Star spots alter the profile of spectral lines, as well known from
Doppler imaging studies of rapidly rotating spotted stars (see e.g. 
Rice \cite{rice}; Strassmeier \cite{strassmeier}).
For slowly rotating stars the distortions of line profile are more subtle
but nevertheless sufficient to be detected as spurious RV variations
(Hatzes \cite{hatzes02}).
The RV variations resulting from the presence of surface features 
typically follow the time scales of the rotational period
of the star (a few days for the active stars for which such signal 
is more easily detectable), but for the long term and sparse sampling
typical of planet search surveys no clear periodicities are often present,
because of the limited lifetime of such features.
RV variations caused by star spots are usually correlated to photometric
variations (Paulson et al.~\cite{paulson04}; Queloz et al.~\cite{hd166435}).
The second contribution is represented by plages, that cause a change of 
the shape of spectral lines, mostly because
of the alteration of the granulation pattern
(Saar \cite{saar03}; K\"urster et al.~\cite{barnard}).
The variations of the area covered by plages along the
magnetic cycle and/or the rotational period then cause RVs variations,
correlated with chromospheric emission.
For some low activity stars, for which the effects of rotational 
modulations are lower than those of the long term magnetic cycle, a fairly 
good correlation between RVs and 
chromospheric emission can be found (Saar \& Fischer \cite{saarfischer}).

In order to disentangle the origin of RV variations 
(keplerian orbital motion {\it vs} activity jitter)
basically three approaches can be pursued.
\begin{enumerate}
\item
To directly search for the presence of  distortions of line profiles 
(ideally on the same spectra on which RVs are derived), as done by 
e.g. Hatzes et al.~(\cite{hatzes98}) for 51 Peg and 
Queloz et al.~(\cite{hd166435}) for HD~166435
\item
To obtain measurements of stellar
activity (chromospheric emission, photometry), possibly simultaneous to RVs, 
to search for correlation between RVs and activity.
Note that one single activity diagnostics may not be enough since
photometry and chromospheric emission are mostly sensitive to
different components of magnetic activity.
\item
To continue the RV monitoring of the object.
In fact, stellar activity typically is not stable on yearly 
timescales, so that an activity signal is not expected to maintain
the same phase and amplitude with time (see Queloz et al.~\cite{hd166435})
\end{enumerate}

For our candidate we followed the second and third approaches,
collecting new RV data and considering literature determinations 
of the chromospheric emission of the star (Wright et al.~\cite{rhk_keck}), 
published after Paper I, to study the evolution of its activity level.
The line bisector variations corresponding to the observed RV variations
are below our detectability threshold, as discussed in Paper I.

%__________________________________________________________________

\section{Radial velocities}

\begin{figure}
   \includegraphics[width=9cm]{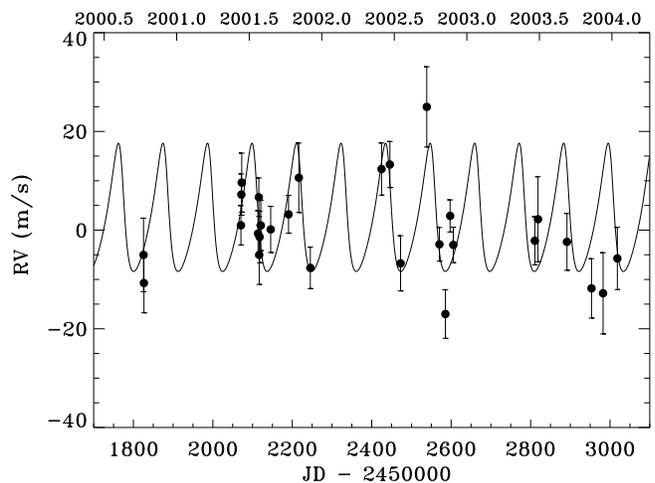}
      \caption{Radial velocity curve for HD~219542~B. The data taken in
        the 2003 season do not follow the tentative orbital solution
        derived in Paper I (overplotted as a solid line)}
         \label{f:hd219542b_rv}
   \end{figure}

The new RVs have been obtained from SARG spectra in the same way as in Paper
I, using the AUSTRAL code (Endl et al.~\cite{austral}).
Six new spectra were acquired from June 2003 to January 2004, hereafter
referred as 2003 season.
Table \ref{t:hd219542b_rv} and Fig.~\ref{f:hd219542b_rv} present the 
full radial velocity data set for HD~219542~B (nightly 
averages)\footnote{For the 2000-2002 velocities, there are minor differences 
with respect to those published in Paper I. These are due to the fact
that some trends in the radial velocity of the $\sim 100$ pixel long chunks
along a spectral order, due to  errors in the wavelength calibration of 
the stellar template (without iodine lines),
are removed  considering the average velocity of each chunk
for the whole dataset. 
This correction then slightly changes with the addition of the new data.
The differences are much smaller than internal errors.
The inclusion of the new data also changes the
normalization.}.
As can be seen in Fig.~\ref{f:hd219542b_rv}, the data taken during
the 2003 season do not follow the tentative orbital solution
derived in Paper I.

\begin{table}
   \caption[]{Differential radial velocities for HD~219542~B}
     \label{t:hd219542b_rv}
      
       \begin{tabular}{lrrclrr}
         \hline
         \noalign{\smallskip}
         JD-2450000 & RV &  err & &  JD-2450000 & RV &  err \\
                    & m/s & m/s & &             & m/s & m/s \\ 
        \noalign{\smallskip}
         \hline
         \noalign{\smallskip}

       1825.51 &    $-$5.0   &    7.4  && 2424.71 &      12.4   &    5.3  \\
       1826.49 &   $-$10.7   &    6.0  && 2445.71 &      13.3   &    4.7  \\
       2070.71 &       1.0   &    4.0  && 2472.70 &    $-$6.7   &    5.6  \\
       2071.71 &       7.2   &    4.2  && 2538.51 &      25.0   &    8.1  \\
       2072.69 &       9.6   &    6.0  && 2570.42 &    $-$2.9   &    3.4  \\
       2113.71 &    $-$0.7   &    4.6  && 2585.45 &   $-$16.0   &    4.9  \\
       2115.69 &       6.7   &    3.9  && 2597.33 &       2.9   &    3.3  \\
       2116.68 &    $-$5.0   &    6.0  && 2605.36 &    $-$3.0   &    3.6  \\
       2117.71 &    $-$1.4   &    5.2  && 2810.69 &    $-$2.2   &    4.9  \\
       2120.72 &       0.9   &    5.1  && 2818.66 &       2.2   &    8.6  \\
       2145.66 &       0.1   &    4.7  && 2891.55 &    $-$2.4   &    5.8  \\
       2190.54 &       3.2   &    3.8  && 2953.35 &   $-$11.8   &    6.0  \\
       2216.47 &      10.6   &    7.0  && 2982.39 &   $-$12.8   &    8.2  \\
       2245.43 &    $-$7.7   &    4.2  && 3018.35 &    $-$5.8   &    6.3  \\

         \noalign{\smallskip}
         \hline
      \end{tabular}
\end{table}

\begin{figure}
   \includegraphics[width=6cm,angle=-90]{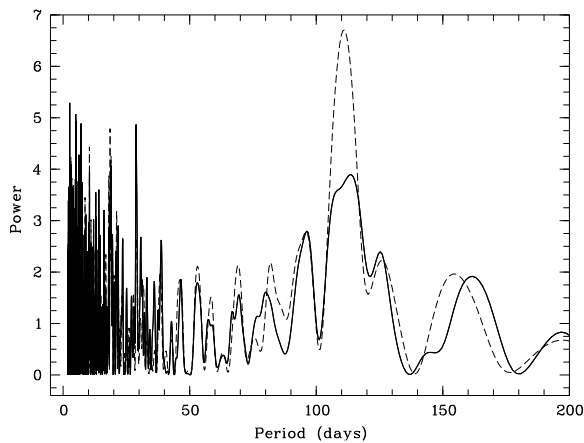}
      \caption{Lomb-Scargle periodogram of radial velocities for HD~219542~B.
        The solid line refers to the full data set while dashed line
        is the periodogram of the data included in Paper I. 
        The inclusion of 2003 data causes a strong decrease of the 112 days
        peak. Note that for display purposes only the period range
        2 to 200 days is shown.}
         \label{f:hd219542b_periods}
   \end{figure}

Fig.~\ref{f:hd219542b_periods} shows the Lomb-Scargle periodogram 
(Lomb~\cite{lomb}; Scargle~\cite{scargle}) for the
data included in Paper I and for the whole dataset presented here.
The power at 112 days sharply decreases with the addition
of the data of the last observing season.

%______________________________________________________________

\section{Chromospheric activity}

Wright et al.~(\cite{rhk_keck})
recently published the measurement of Ca II H \& K emission for the whole
samples of  the Keck and Lick planets searches.
HD~219542~A and B were recently added to the Keck planet search sample.
Overall, 11 and 15 measurements from June 2002 to July 2003 are presented
for the two stars respectively.
The mean {\em S} index for HD 219542~A and B result of
0.158 and 0.204 respectively. The calibration of 
Noyes et al.~(\cite{noyes94})
coupled with the colors of the stars as given by Simbad, i.e. 
{\em (B-V)}=0.64 and 
0.69 for HD~219542A and B respectively\footnote{The  colors and absolute 
magnitudes adopted by Wright et al.
($\Delta$(B-V)=0.014 mag; HD~219542~B brighter by 0.59 mag), indicate 
that they probably took
the magnitude and colors of HD~219542~A from the Simbad object HD~219542
and those of HD~219542~B from the Simbad object CCDM~J23166-0135AB, which
is actually the composite HD~219542~A+B, instead that for BD~-02~5917B.}
gives $\log R^{'}_{HK}=-5.01$ and $-4.80$.
Our measurement of Paper I was
based on a single FEROS spectrum and has a fairly large error because of the
low signal and the paucity of available calibrators with low activity level.

\begin{figure}
   \includegraphics[width=9cm]{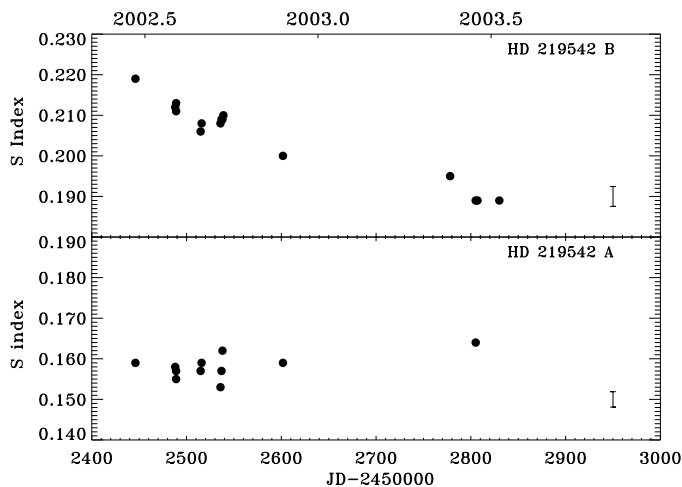}
      \caption{Ca II H \& K emission measurements of HD~219542~A (lower 
               panel) and  HD~219542~B (upper panel)  from Keck 
               (Wright et al.~\cite{rhk_keck}). Typical error bars on 
               differential values of S index measured at Keck are shown 
               in the right part of each panel. These results show an 
               enhanced activity in 2002 and a
               sharp decline over the following year for HD~219542~B.
               The activity level of 
               component A is fairly constant and lower than component B.}
         \label{f:rhk}
   \end{figure}

Single epoch measurements are also included in Wright et al.~\cite{rhk_keck}
(available at astro.berkeley.edu/$\sim$jtwright/CaIIdata/tab1.tex). 
These were obtained using the differential technique described in their 
paper, reaching errros of 1.2\%.
They are plotted in Fig.~\ref{f:rhk}.

\section{Discussion}

The differential Ca II H \& K measurements by Wright et al.~(\cite{rhk_keck})
indicate a much higher activity level for HD~219542~B 
during the 2002 season with respect to 2003. 
The activity level of HD~219542~A is instead lower and nearly constant,
with a possible small increase in the 2003 season.
When considering the radial velocity curve, it should be noticed that
the dispersion of RVs of HD~219542~B is larger during the 2002 season.
Table \ref{t:season} shows the seasonal mean and dispersion of RVs and Ca II
H \& K 
emission, the measured RV dispersion excess (calculated as the quadratic 
difference between the observed dispersion and the mean internal errors) 
as well as the radial velocity jitter expected on the basis of the 
calibration by Marcy (2002, private communication).

\begin{table*}
   \caption[]{Seasonal mean and dispersion of radial velocities 
              and Ca H\&K emission for HD~219542~B}
     \label{t:season}
      
       \begin{tabular}{lrrrrrrr}
         \hline
         \noalign{\smallskip}
         Season & $N_{data}$ & Mean RV &  rms RVs & err & $\log R^{'}_{HK}$ & obs jitter & 
         exp jitter\\
                & &  m/s &  m/s & m/s &  & m/s & m/s \\

         \noalign{\smallskip}
         \hline
         \noalign{\smallskip}
                                                                               
2001 & 12 &    2.0$\pm$1.6 &  5.6 & 5.0 &   --    &  2.6 & -- \\
2002 &  8 &    3.0$\pm$4.7 & 13.3 & 5.1 & $-$4.78 & 10.1 & 7.1 \\
2003 &  6 & $-$5.4$\pm$2.6 &  6.5 & 6.8 & $-$4.86 &  0.0 & 6.1 \\
all  & 28 &    0.0$\pm$1.7 &  9.0 & 5.6 & $-$4.80 &  7.1 & 6.8 \\

         \noalign{\smallskip}
         \hline
      \end{tabular}
\end{table*}

These results strongly suggest that HD~219542~B underwent 
a phase of enhanced activity during 2002, while the activity
decreased significantly
in 2003. On the basis of the low RV scatter, the activity
level could have been low also in June-July 2001.

We note that the seasonal variations of rms RV dispersion are much
larger than those predicted by the jitter calibration. The
global scatter on longer timescale (4 years) is instead much
better predicted by such relation.
This indicates that the estimates on RV jitter based on the activity
level of the star should be more reliable when considering a fairly
long time coverage,
while on short time scales the calibration might severely underestimate 
or overestimate the actual jitter. This fact may be explained considering 
that during the phases of higher activity the appearance of 
spots and plages causes short term RV (and photometric) variations
according to the rotational phases (see e.g. 
Paulson et al.~\cite{paulson04}),
while the absolute value of the chromospheric 
emission might be only moderately affected. 

It appears that the high activity phase of HD~219542~B during 2002
was relatively short lived when compared to that of the Sun.
The star might have a fairly short and/or irregular activity cycle.
Alternatively, its activity behaviour might be similar 
to that of 15 Sge, that has a cycle with activity enhancements  
lasting about 1 year every 17 years
(Baliunas et al.~\cite{baliunas}, Wright et al.~\cite{rhk_keck})
and low activity level at other epochs. 

HD~219542~A  instead shows no sign of enhanced or variable activity. 
The age of the two components derived using the relation by 
Donahue (\cite{donahue}) is 5.7 and 2.7 Gyr for HD~219542~A and B 
respectively. This age difference is larger than typically obtained
in coeval systems (Henry et al.~\cite{henry96}) but
similar to that derived for the Sun when using
the maximum and minimum {\em S} values along its magnetic cycle.

In Paper I we noted that the 2002 RV data could be fitted quite well
by sinusoidal variations with a period of 18.5 days. Both the
RV and Ca H \& K emission data are not sufficient to firmly establish
the rotational period of the star\footnote{The power at 18.5 days decreased
with the inclusion of 2003 data, see Fig~\ref{f:hd219542b_periods}.}. 
However, a period of 
18.5 days is compatible within the uncertainties to that obtained using the 
Noyes et al.~(\cite{noyes94}) calibration.

\begin{figure}
   \includegraphics[width=6cm,angle=-90]{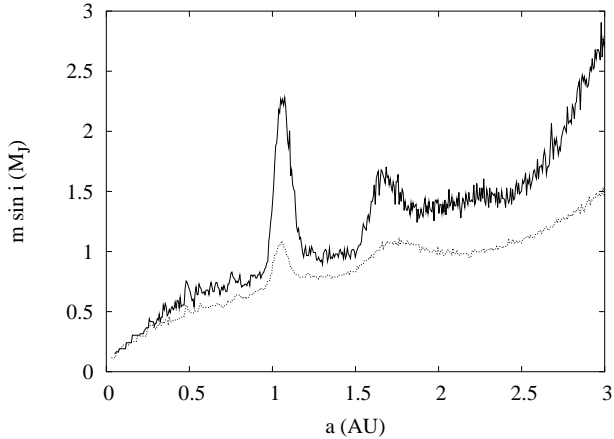}
      \caption{Limits on planets around HD~219542~B in circular orbits
                (dotted line) and eccentric orbits (solid line).}
         \label{f:limits}
   \end{figure}

In spite of the moderate activity level of the star our RVs are
anyway of some usefulness for planet search.
We derived the upper limits on the planets which are still 
compatible with our data (Fig.~\ref{f:limits}). 
We use the same technique developed in Paper I, that
allows the determinations of the upper limits on the mass of 
companions in eccentric orbits,
while most of the upper limits determinations in literature 
consider only circular orbits.
With the exclusion of a small window around 1 year period, planets
with $m \sin i > 1 M_{J}$ can be excluded within 1.5 AU.
The peak around 1 year can be explained considering that a planet with period 
close to 1 year, fairly high eccentricity, and longitude of periastron and 
orbital phases in  suitable ranges would cause a radial velocity curve 
nearly flat most of the time with  RV variations concentrated
on the times on which the target is not observable.

%______________________________________________________________

\section{Conclusion}

The continuation of the radial velocity monitoring and the
multi-epoch measurements of the Ca II H\&K emission indicate that the
low amplitude RV variations of HD~219542~B presented in Paper I are 
likely due to stellar activity. This star should therefore be removed from
the list of extrasolar planet host stars. 

The available data suggest that the star underwent a relatively short-lived 
phase of enhanced activity during the 2002 season. 

This study confirms the relevance
of the activity-related phenomena in the RV planet searches
and, on the other hand, the great impact of the high precision
Doppler surveys in improving our understanding
of the stellar activity cycles. Even for stars with modest
activity level, the discovery of planets with
amplitude of 10-15 m/s requires a long term monitoring
to check for the stability of the signal as well as ancillary
measurements of activity indicators (see e.g. Hatzes et al.~\cite{epseri}).

\begin{acknowledgements}

We thank the TNG staff for its help in the observations.
We are grateful to G. Marcy for providing us its jitter calibration. 
We thank the referee, Dr. R. Freire-Ferrero, for his stimulating comments.
This research has made use of the SIMBAD database, operated at CDS, 
Strasbourg, France. 
ME is supported by NASA Grant NAG5-13206

\end{acknowledgements}

\end{document}